\PassOptionsToPackage{table}{xcolor}
\documentclass[sigconf]{acmart}
\usepackage{makecell} 
\usepackage{xcolor} 
\usepackage{fancyvrb} 
\usepackage{multirow} 
\AtBeginDocument{%
  }

\copyrightyear{2024}
\acmYear{2024}
\setcopyright{acmlicensed}\acmConference[Websci '24]{ACM Web Science Conference}{May 21--24, 2024}{Stuttgart, Germany}
\acmBooktitle{ACM Web Science Conference (Websci '24), May 21--24,
2024, Stuttgart, Germany}
\acmDOI{10.1145/3614419.3644018}
\acmISBN{979-8-4007-0334-8/24/05}

\acmSubmissionID{123-A56-BU3}

\input cyracc.def
\font\tencyr=wncyr10
\def\cyr{\tencyr\cyracc}

\begin{document}

\title{Improved methodology for longitudinal Web analytics using Common Crawl}

\author{Henry S. Thompson}
\orcid{0001-5490-1347}
\affiliation{%
  \institution{The University of Edinburgh}
  \city{Edinburgh}
  \country{United Kingdom}
}

\renewcommand{\shortauthors}{Thompson}

\begin{abstract}
  Common Crawl is a multi-petabyte longitudinal dataset containing
  over 100 billion web pages which is widely used as a source of
  language data for sequence model training and in web science
  research.  Each of its constituent archives is on the order of 75TB
  in size.  Using it for research, particularly longitudinal studies,
  which necessarily involve multiple archives, is therefore very
  expensive in terms of compute time and storage space and/or web
  bandwidth. Two new methods for mitigating this problem are presented
  here, based on exploiting and extending the much smaller (<200
  gigabytes (GB)
  compressed) {\em index} which is available for each archive.  By
  adding Last-Modified timestamps to the index we enable longitudinal
  exploration using only a single archive.  By comparing the
  distribution of index features for each of the 100 segments into
  which archive is divided with their distribution over the whole
  archive, we have identified the least and most representative
  segments for a number of recent archives.  Using this allows the
  segment(s) that are most representative of an archive to be used as
  proxies for the whole.  We illustrate this approach in an analysis of
  changes in URI length over time, leading to an unanticipated insight
  into the how the creation of Web pages has changed over time.
\end{abstract}




\maketitle

\section{Introduction}

Common Crawl \cite{common_crawl} is a very-large-scale corpus
containing petabytes of data from more than 100
archives.  It contains over 100 billion web pages, more than 99\% of which
are HTML formatted, collected since 2008. It is hosted online free of charge
courtesy of Amazon Web Services’ Open Data Sponsorship program \cite{AWSData}.

Existing Web Analytics literature focuses on small subsets of the Web,
or only examines the distribution of Top-Level Domains (TLDs).
Although Common Crawl offers a large and freely available source of
data, it has rarely been used as a basis for studying the evolution of
the Web.

One obvious reason for this is the considerable cost, both in compute
time, web bandwidth and storage that is involved in processing multiple
archives.  The average size of a compressed archive has grown from
50 terabytes (TB) in 2019 to 100TB in 2023.

In this work we explore how to take advantage of two aspects of
Common Crawl archives to mitigate this problem:

\begin{enumerate}
\item Each archive is packaged in 100 randomised subsets called
{\bf{\em segments}} of approximately equal size;
\item Each archive since 2013 comes with a sharded
index \cite{cci}, which provides descriptive metadata
for every retrieved item therein.
\end{enumerate}

The index for a given archive is available for separate download and
is less than 200GB in size.  It is thus possible to explore the
evolution of a number of important aspects of Web architecture which
are manifest in the index with much less computational cost compared
to that required for working with the corresponding complete archives
themselves.

Chen \cite{cookies} suggested that \textit{if} one could guarantee
that "each segment [of a Common Crawl archive] has similar
distribution [to the whole]", \textit{then} processing only a single
segment per year would provide comparable results to those for the
whole, at 1\% of the cost.  We take up that suggestion here by using
the index to measure the extent to which each segment is
representative of the whole archive, we can use the segment(s) that
are most representative of an archive as proxies for the whole, again
reducing the computational cost involved even when the properties of
interest are \textit{not} available from the index.

In what follows we report on a two-part study.  The first explores the
use of index features from four archives from
August/September of 2019, 2020, 2021 and 2023 to determine how we can best
measure segment representativeness in order to chose proxies.  The second
uses this approach to evaluate the hypothesis that the length of URIs
is growing, and that this growth is due primarily to a growth in the
query URI component.  We conclude with some surprising evidence for
the ongoing change in the way web pages are created, from human
authoring to automatic generation.

\section{Materials}
\begin{quote}
  ``The Common Crawl corpus contains petabytes of data collected since
  2008. It contains raw web page data, extracted metadata and text
  extractions.

The Common Crawl dataset lives on Amazon S3 as part of the Amazon Web
Services’ Open Data Sponsorships program. You can download the files
entirely free using HTTP(S) or S3.'' \cite{ccf1}
\end{quote}

Each Common Crawl archive has six main components \cite{ccf2}:

\begin{enumerate}
\item Data files
\begin{enumerate}
\item Successful retrievals
\begin{enumerate}
\item {\bf WARC} files: Complete HTTP request and response; archive metadata
\item {\bf WAT} files: Response metadata only
\item {\bf WET} files: Text content (HTML responses only)
\end{enumerate}
\item Unsuccessful retrievals
\begin{enumerate}
\item {\bf robots.txt} files: HTTP headers and message body
\item {\bf non-200 responses} files: HTTP headers and message body for
  e.g. not found, server fault, redirect
\end{enumerate} 
\end{enumerate}
\item {\bf URI index} files: 300 shards, alphabetical by URI inverse
  hostname, plus binary-searchable master index
\end{enumerate}

All the constituent files are compressed, with the exception of the
URI master index.

The data files are divided into 100 segments, numbered from 0 to 99,
with equal numbers of files, but not necessarily of
pages\footnote{Strictly speaking the contents of archives are not
  pages, but records of HTTP requests and responses. For
  successful requests the response body is a (text/html or
  application/pdf or ...) representation of one or more
  pages, and we'll use \textit{page} below in this informal way.}.

The details of the four Common Crawl archives we used are given in
\renewcommand\tableautorefname{Tables}
\autoref{tab:ccdata1}
\renewcommand\tableautorefname{and}
\autoref{tab:ccdata2}\renewcommand\tableautorefname{Table}.  The two-digit number
at the end of the Archive ID gives the week in which the archive was
carried out.  Each gzipped WARC-format \cite{warc} file in these
archives contains around 50,000 records of successful HTTP(S)
request-response exchanges. The CDX-format \cite{cdx} index of request
URIs provides file-ids, offset and length for all records in the WARC,
robots.txt and non-200 response components (details below).

\begin{table}
  \caption{Scale of archives used  }
  \label{tab:ccdata1}
  \begin{tabular}{llll}
    Archive ID & WARC files & WARC Size (compressed) \\
    \midrule
    CC-MAIN-2019-35 & 56,000 & 54TB \\
    CC-MAIN-2029-34 & 60,000 & 49TB \\
    CC-MAIN-2021-31 & 72,000 & 75TB \\
    CC-MAIN-2023-40 & 90,000 & 98TB \\ \bottomrule
  \end{tabular}
\end{table}

\begin{table}
  \caption{Size of archive components (in millions of retrievals) }
  \label{tab:ccdata2}
  \begin{tabular}{r|rrrr}
 & 2019-35 & 2020-34 & 2021-31 & 2023-40 \\
    \midrule
WARC & 2,955 & 2,450 & 3,165 & 3,445 \\
non-200 & 549 & 520 & 465 & 553 \\
robots.txt & 114 & 108 & 86 & 90 \\
 \midrule
Total & 3,618 & 3,078 & 3,716 & 4,089 \\ \bottomrule
  \end{tabular}
\end{table}
\subsection{URL index}
\begin{quote}
  ``The index format is relatively simple: It consists of a compressed
  plaintext index (with one line for each entry) compressed into
  gzipped chunks, and a [master] index of the compressed chunks. This
  index is often called the ‘ZipNum’ CDX format \cite{cdx} and it is the same
  format that is used by the Wayback Machine at the Internet Archive.'' \cite{cci}
\end{quote}

Taken together, the two parts of the URI index make it possible to
access a single request-response record, either from a local copy of
the WARC component\footnote{As noted earlier, the index actually
  covers the robots.txt and non-200 responses components as well, but
  the work reported here makes no use of that.} or by via an HTTP
request.

An abbreviated example is the best way to explain how this works.
Suppose we want to see if the XML specification \cite{xml}, whose URI
is {\tt https://www.w3.org/TR/xml/}, is in the August 2021 archive, for
which we have the index.

Both the master index ({\tt cluster.idx}) and the individual
primary index files are keyed with a {\em urlkey}, which is a version of a
URI, canonicalised using a version of the Internet Archive's {\em
  Sort-friendly URI Reordering Transform} \cite{SURT}\footnote{Actual
  implementations not only do more than what is described here, they
  also differ with respect to the how they handle various corner
  cases, {\em caveat emptor}.}:

\begin{itemize}
\item Remove {\tt http(s)://};
\item replace A-Z with a-z throughout;
\item if the authority component begins with {\tt www.}, remove it;
\item reverse the order of the authority component, replace any
  periods with commas and insert a right-parenthesis at the end;
\item if the path ends with a slash, remove it.
\end{itemize}

For our example, this would turn {\tt https://www.w3.org/TR/xml/} into
{\tt org,w3)/tr/xml}.

The lines of the primary and master index files consist of
space-separated fields, the first of which is a urlkey, and the lines
are in alphabetical order based on that field.

There are 300 primary index files, and each file is compressed into
blocks of 3000 lines.  The master index file contains one line for
each of those compressed blocks, keyed by the urlkey for the first line
in the block.  The remaining fields in the master index give the
primary index file name
and the offset and length of the block in that file.

This provides for very efficient search for a URI, as follows:
\begin{enumerate}
\item Convert to a urlkey;
\item binary search in the master index for the last line whose key
  is less than or
  equal to that urlkey, and note the primary index file name, offset and length therein;
\item extract the block of the given length at the given offset {\em from
  the gzipped version} of the given primary index file and unzip it;
\item binary search therein for the last line whose key is less than or
  equal to your urlkey:  if equal, you win, if not, you lose.
\end{enumerate}

Note that this depends on an important property of the gzip compression
format \cite{gzip}, which is used by Common Crawl for compression: a
single compressed file may consist of the concatenation of separately
compressed files.  Not only the primary index files, but also the WARC
files themselves exploit this.  For each successful retrieval there is a
separate gzipped block containing the three WARC-format records for
HTTP request, HTTP response and archive metadata.  This allows
individual entries in the primary index file to reference individual
retrievals by file ID, offset and length in a gzipped WARC file.

Although there are only on the order of $3{\times}10^9$ successful retrievals in
recent archives, because the index \textit{also} covers the {\bf robots.txt} and
{\bf non-200} retrievals, there are around $3.6{\times}10^9$ index
entries in total, split into $300$ primary index files.  Each individual index file
thus has on the order of $1.2{\times}10^7$ lines.  This in turn
means that, since the compressed blocks in each file contain $3,000$
lines, a single primary index file is split into about $4,000$ blocks and the
master index, with one line per block, has about 1.2 million lines.

Looking up a given URI thus takes around 21 tests in the master index
(step 2 above) and 12 tests in the resulting primary index block (step
4) and only requires unzipping one $3,000$-line block.

Note that if you win, there may be more than one entry in the primary
index which matches, as we will see with our example, which goes like
this:

\begin{enumerate}
\item Convert {\tt https://www.w3.org/TR/xml/}\newline{}into urlkey: {\tt org,w3)/tr/xml}
\item In the primary index, we find on lines 843315 and 843316:
\begin{small}
\begin{verbatim}
org,w3)/tr/tr.xml cdx-00253.gz 557238519 185309
org,w3)/wai/videos/standards-and-benefits/ja
 cdx-00253.gz 557423828 182738
\end{verbatim}
\end{small}
\item Extract 185309 bytes from cdx-00253.gz beginning at offset
  557238519 and unzip the resulting block of 3000 lines
\item Binary search therein finds the following lines, beginning at line 1518:
\begin{small}
\begin{Verbatim}[commandchars=\#\[\]]
org,w3)/tr/xml
 20210613173657
 {"url": "https://www.w3.org/TR/XML/",
 "mime": "text/html",
 "mime-detected": "text/html",
 "status": "301",
 "digest": "LQRWZ7SMYYGCL55UJSVAS3BY64YNZ4DQ",
 "length": "743",
 "offset": "27241472",
 "filename": "crawl-data/CC-MAIN-2021-25/segments/\
              1623487610196.46/#vbf[#normalsize[crawldiagnostics]]/\
              CC-MAIN-20210613161945-20210613191945\
             -00275.warc.gz",
 "redirect": "https://www.w3.org/TR/xml/"}
org,w3)/tr/xml
 20210613173657
 {"url": "https://www.w3.org/TR/xml/",
 "mime": "text/html",
 "mime-detected": "application/xhtml+xml",
 "status": "200",
 "digest": "AOMNGHUQLUKLHHWBNUL7MOVXKIUX522W",
 "length": "55091",
 "offset": "968583998",
 "filename": "crawl-data/CC-MAIN-2021-25/segments/\
              1623487610196.46/#vbf[#normalsize[warc]]/CC-MAIN-\
              20210613161945-20210613191945-00371.warc.gz",
 "charset": "UTF-8",
 "languages": "eng"}
\end{Verbatim}
\end{small}
\end{enumerate}

The first of these lines is the entry for {\tt
  https://www.w3.org/TR/XML/}, which resulted in a redirection (status
code 301) when retrieved, and the record of this is therefore
contained in the non-200 archive component, indicated by the
``crawldiagnostics'' part of the filename.

The second line, with a status code of 200, is the one we were looking
for, with ``warc'' in its filename.

The format of the primary index lines is simple:
\begin{verbatim}
      urlkey<space>timestamp<space>JSON-array
\end{verbatim}

In addition to the WARC filename, offset and length which allow direct
access to the complete response, the  JSON-array always contains the
following features:
\begin{itemize}
\item ``url'': as contained in the successful HTTP request
\item ``status'': HTTP response status code 
\item ``mime'': from the {\tt Content-Type} HTTP response header
\item ``digest'': a SHA1 hash of
the response payload.
\end{itemize}

HTML responses have two additional computed features:

\begin{itemize}
\item ``charset'': from the {\tt Content-Type} HTTP response header
\item  ``mime-detected'': 'sniffed' from the response body itself,
using Apache Tika \cite{tika}
\item ``languages'': computed from the HTML using
CLD2 \cite{cld2}.  May be absent, or
contain up to three, comma-separated, lower-case ISO three-letter
language codes.
\end{itemize}

\section{Background}\label{background}
There are relatively few reports of large-scale longitudinal studies
of the Web based on data from Common Crawl.  As was the case with our
own earlier work \cite{TandT2018}, which was based on only two Common
Crawl archives, such studies often draw on only a small number of
archives.  This is likely to be at least in part because of the large
amount of processing required to tabulate some phenomenon from a
number of multi-TB archives.

For example although Chapuis et al.
\cite{Chap2020} used seven archives from between 2016 and 2019, they
commented that ``Parsing the content of all the webpages contained in
a snapshot [archive] of CC is time consuming'', so they proceeded to
extract only 1\% of those seven archives for use in most of their
analysis.  They don't discuss the question of the mechanism by which
that 1\% was chose or how representative it was, but it does appear
from their code archive that it was a single segment.  Similarly,
Luciano and Viviani \cite{LandV2021} say ``[B]oth downloading and analyzing
the Common Crawl are time-consuming and costly endeavors'' and go on
to sample only 1\% of a single archive.

In another recent example, a study of HTML validity \cite{FandS2022},
Florian and Stock appear to have used Common Crawl indices to identify
candidate pages for analysis, but only actually downloaded around 1.8
million pages per archive from one archive per year from 2015 through
2022, that is, less than .1\% of each.

Other work, such as \cite{EberEtAl2015}, \cite{PanchEtAl2018} and
\cite{ChiniahEtAl2019}, uses only (parts of) single Common Crawl
archives or even, as in \cite{DuEtAl2017}, despite actually being
focused on testing representativeness, don't specify which or how
many Common Crawl archives are used, only the (small) number of actual
pages processed or analysed.

I'm have not been able to find any previous work which explicitly
addresses the particular issues that concern us here, namely measuring
the representativeness of individual segments and using Last-Modified
headers to see further into the past.

On the general issue of Common Crawl's goals, policies and use in
the recent explosion of Large Language Models, a recent research paper
from Mozilla \cite{moz2024} is the most comprehensive analysis I've
seen.

\section{Part 1: Measuring and exploiting segment representativeness}
\subsection{Methodology}
\subsubsection{Analysing archive properties}
Our first hypothesis is that by measuring the correlation between the
distribution of features of the individual segments in an archive and
their distribution in the archive as a whole, we can obtain a measure
of the extent to which each segment in a Common Crawl archive contains
a representative sample of the whole of that archive.

To test this we would like an easily available but rich property of
the data whose overall distribution can be compared to its
distribution in each segment.

Features found in the index satisfy the easily available
requirement.  We start by testing the use of a composite
property based on the ``mime'' and ``mime-detected'' features.

For each of the four archives, we first tabulated the frequency of
such pairs from the index entries for all successful retrievals,
simplifying slightly by replacing a ``mime-detected'' value identical
to the ``mime'' value with {\bf ditto}.  We did this for each segment
first, and then merged the results to get a tabulation for the
complete archives.

For example, \autoref{tab:mimesample} shows the top 10 pairs from the 2019 archive.

\begin{table}
  \caption{Sample from whole-archive mime tabulation from 2019-35}
  \label{tab:mimesample}
  \begin{tabular}{rll}
    frequency&mime&mime-detected\\
\midrule
2,232,464,436&text/html&ditto \\
650,577,285&text/html&application/xhtml+xml \\
40,022,222&unk&text/html \\
3,985,789&application/atom+xml&ditto \\
3,879,977&application/pdf&ditto \\
3,741,189&image/jpeg&ditto \\
2,741,054&unk&application/xhtml+xml \\
2,488,581&application/rss+xml&ditto \\
1,565,481&text/xml&application/rss+xml \\
1,229,831&text/plain&ditto \\ \bottomrule
    \end{tabular}
  \end{table}

For each archive, we then created a merged tabulation for the top 100
pairs in that archive, with the total count and the count from each
segment.  \autoref{tab:mimemerged} shows a small extract from this
101 x 101 table for 2019:

\renewcommand\cellalign{rc}
\begin{table}
  \caption{Example counts from merged mime tabulation for 2019-35}
  \label{tab:mimemerged}
  \begin{tabular}{r|rr|rrr}
&\multicolumn{2}{c}{whole archive}&\multicolumn{3}{c}{segment counts} \\
mime pair&rank&count&71&72&73 \\
\midrule
\rowcolor{gray!15}
text/plain application/mbox&52&37711&435&364&397 \\
\makecell{application/octet-stream \\
             application/x-tika-msoffice}&53&37414&{\bf\textcolor{red}{354}}&nan&{\bf\textcolor{red}{2}} \\
   \rowcolor{gray!15}
\makecell{application/octet-stream \\ text/x-log}&54&36352&{\bf\textcolor{red}{651}}&248&{\bf\textcolor{red}{345}} \\ \bottomrule
\end{tabular}
\end{table}

Two aspects of this example deserve comment:
\begin{enumerate}
\item \textbf{The `nan' in the middle} This means there was some data
  missing. Since we used the top 100 pairs from the whole corpus when
  taking counts from the individual segments, there are a few cases
  where a given pair from the whole-archive-top-100 did not occur at
  all in a given segment.  Such drop-outs are recorded as `nan' (not a
  number).  The cutoff at 100 was chosen to keep this
  from happening very often, so in fact the four years had only 6, 5,
  4 and 1 cases of this respectively out of $10,000$, and the impact of
  the missing entries is negligible.
\item \textbf{The red, bold-face pairs of counts}  These highlight mis-matches between the
ranks of 53 and 54 in the whole
archive for the 2nd and 3rd media pairs and their counts in segments
71 and 73.  

\end{enumerate}

Those two mis-matches suggest a way to measure the representativeness of the
individual segments:  the rank correlation between the columns of this
table.  We use rank correlation because, as is already evident from
\autoref{tab:mimesample}, the
distributions involved are far from normal.

We used the {\tt stats.spearmanr} function in the Python
Scipy library \cite{scipy} to compute a full 101 x 101 array of the rank
correlation between every possible pair in the 100 x 101 tabulation of
counts, using the 'omit' policy for dealing with missing (`nan') cells.

\autoref{tab:res1} shows a small extract, using the same columns as
in \autoref{tab:mimemerged}, from the resulting correlation\footnote{We'll just say ``correlation'' from now on, but
  all numbers described as such are actually {\em rank} correlations} array
for 2019.
For what it's worth, the p-values produced by the {\tt
  stats.spearmanr} function were tiny, on the order of $10^{-60}$ or
less.  In \autoref{represent} below we'll provide confidence intervals, which
give a better sense of the reliability of the correlation values.

\begin{table}
  \caption{Example rank correlations}
  \label{tab:res1}
  \begin{tabular}{r|rrrr}
&whole&\multicolumn{3}{c}{segments} \\
&archive&71&72&73 \\
\midrule
whole archive&1.& 0.947& 0.949& 0.937 \\
segment 71&0.947& 1.&0.896&0.894 \\
segment 72&0.949& 0.896&1.&0.899 \\
segment 73&0.937& 0.894&0.899&1. \\ \bottomrule
 \end{tabular}
\end{table}

The results are necessarily symmetrical around the self-correlation
diagonal.  The pattern shown in this example, where the correlation
between each segment and the whole is noticeably higher than the
cross-correlations with any other segment, is repeated throughout the
whole tabulation.

\subsubsection{Using one property as a proxy for another}\label{proxy}
Our goal is to explore the possibility of using a property we have
complete data for to predict the best segment(s) to use in place of
the whole.  This may be particularly useful for studying
properties for which we \textit{don't} have complete data.  We will
illustrate this using properties based on two other features from the index for 2019-35:
\begin{enumerate}
\item \textbf{language property}: For \texttt{text/html} responses,
  Common Crawl uses the CLD2 tool to populate the ``languages'' field
  in the index with up to 3 language codes, in rank order, from which
  we used only the first.  As for the mime case, we produced a 101 x 101
  table of rank correlations with respect to the number of occurrences
  of each language in the archive as a whole and in each segment.  We
  included only the top 100 languages in order to
  keep the number of NaNs to a reasonable level.
\item \textbf{length property}: The index includes the (zipped) ``length''
  of the response, which we tabulated as a percentile.  Once again
  this gave us a 101 x 101 table of rank correlations.
\end{enumerate}

For the 2019-35 data, we then tested how successful using one or more
of the best (in terms of correlation with the whole) segment for one
property was, in terms of those same segments' performance on a different property
(again, in terms of correlation with the whole).  We report this as a
percentile \textit{vis-a-vis} the range of values for all 100 segments
(as shown for the mime property in \autoref{tab:res1}).

For example, taking the top five segments in terms of correlation with
the whole for the merged mime property (the \textbf{basis}) and
averaging those segments' correlation with the whole for the language
property (the \textbf{target}) gives $57.7$.  This is better than the
overall mean of $54.6$ for individual segments vs. the whole on that
measure, by about .4 of a standard deviation.  

We tested each of our three measures as the basis against the other
two as target, giving a total of six pairings.  For each of these
pairs we tested each of the top ten basis segments again the target.
We then tabulated the percentile performance of the top one, and
averages of the top two, three, ...ten.

Note that this is only a simulated test of finding proxies for
properties that are \textit{not} in the index.  We did this precisely
because this way we were able to compare our results across a wider
range of data than working with \textit{actual} whole-archive results,
which would have taken more computational resources than we had
available.

\subsection{Results}
\subsubsection{Mime property}\label{represent}
As noted above, we looked at the full 101 x 101 correlation matrix
for the distribution of the mime property, derived from the index, tabulated
for each whole archive and its 100
individual segments.

Although the cross-correlations \textit{between} segments have some interesting
properties, for our purposes we will focus here on the 100
correlations between each segment and the whole archive.
\autoref{tab:seg_stats} tabulates values from Scipy's {\tt stats.describe}
as applied to these for our four archives.

\begin{table}
  \centering
  \caption{Descriptive statistics for segment-vs-whole rank correlations
    between mime property distributions}
  \label{tab:seg_stats}
  \begin{tabular}{r|rrrrr}
    Archive&N&min&max&mean&variance  \\
    \midrule
    2019-35&100&0.898&0.962&0.932&0.0002 \\
    2020-34&100&0.891&0.958&0.929&0.0002   \\
    2021-31&100&0.873&0.965&0.928&0.0004  \\
    2023-40&100&0.853&0.956&0.926&0.0004 \\ \bottomrule
    \end{tabular}
\end{table}

This all looks good from the point of view of our ultimate goal, that
is, to identify segments in any archive that accurately reflect the
archive as a whole, in that the `best' segments
each year have very high correlations (>.95) with the whole archive.

The distribution of the correlations between segments and the whole
archive is reasonably close to normal in all  years, with
Shapiro-Wilks values of 0.972, 0.975, 0.945, 0.84, p<=0.5 in all
cases.

\begin{figure}
  \centering
  \includegraphics[width=\linewidth]{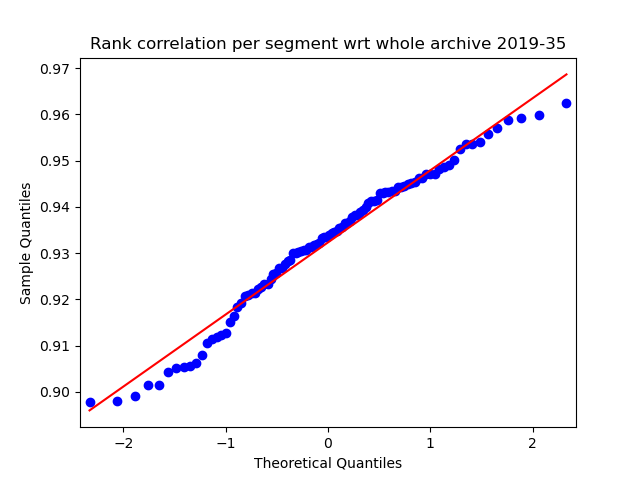}
  \caption[QQ plot for segment-vs-whole correlations, 2019-35
    ]{QQ plot for segment-vs-whole correlations, 2019-35.
     \textnormal{Based on the distribution of the mime property}}
  \label{fig:qqa}
  \Description{Pretty much linear diagonal from bottom-left to
    top-right, some skew towards top and bottom}
  \rule[2ex]{\linewidth}{1pt} %
\end{figure}

The QQ-plot produced by the StatsModels {\bf api.qqplot} \cite{statm}
for the 2019-35 data is shown in \autoref{fig:qqa} and the other
three are similar, although the 2023 plots show 5 outliers at the
bottom end, consistent with the somewhat lower Shapiro-Wilks result.

\begin{figure}
  \centering
  \includegraphics[width=\linewidth]{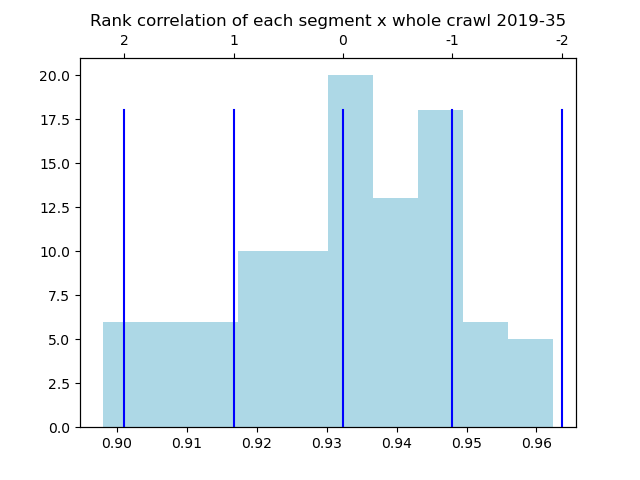}
  \caption[Histogram for segment-vs-whole correlations, 2019-35
  ]{Histogram for segment-vs-whole correlations, 2019-35.
    \textnormal{Based on the distribution of the mime property.  Blue
      vertical lines show standard deviations from the mean.}}
  \label{fig:hist}
  \Description{Approximately normal, with all falling within 2
    standard deviations of the mean}
  \rule[2ex]{\linewidth}{1pt} %
\end{figure}

A histogram with mean and standard deviations marked for 2019-35
is shown in \autoref{fig:hist}.

\begin{figure}
  \centering
  \includegraphics[width=\linewidth]{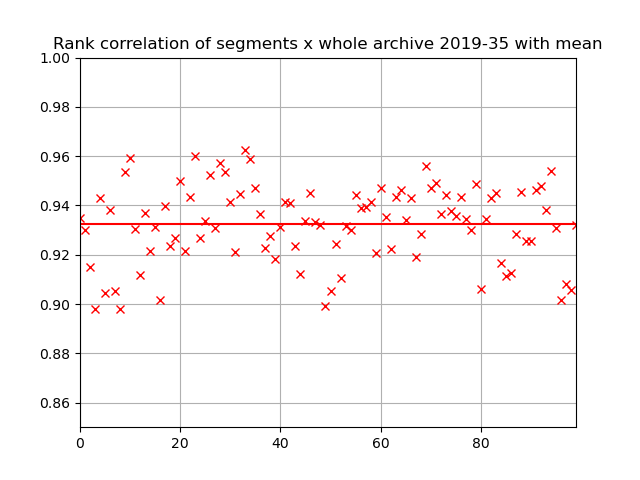}
  \caption[Segment-vs-whole correlations, ordered by segment, 2019-35 ]{Segment-vs-whole correlations, ordered by segment, 2019-35. \textnormal{Based on the distribution of the mime property}}
  \label{fig:plot}
  \Description{Best is a .96 within .945 to .975 range, worst at .90
    within .86 to .94 range}
  \rule[2ex]{\linewidth}{1pt} %
\end{figure}

The same data is plotted per-segment, with the overall mean, in \autoref{fig:plot}.

\begin{figure}
  \centering
  \includegraphics[width=\linewidth]{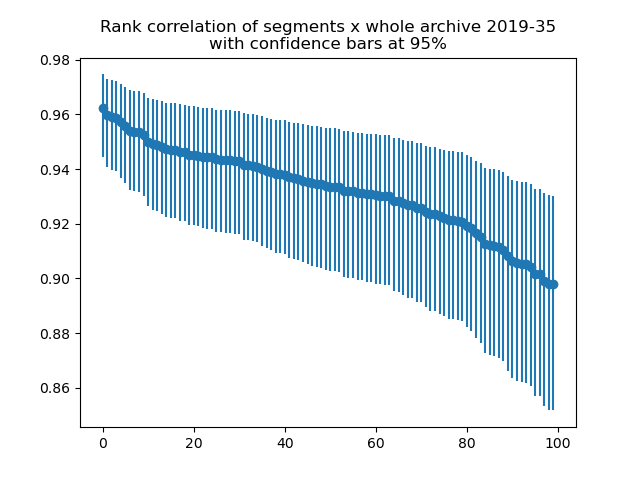}
  \caption[Segment-vs-whole correlations, ordered by correlation, with
  95\% confidence intervals, 2019-35. ]{Segment-vs-whole correlations, ordered by correlation, with
  95\% confidence intervals, 2019-35. \textnormal{Based on
      the distribution of the mime property.  Calculated using the
  atanh approach as described in \cite{atanh}.  The worst error bar is
\textit{just} disjoint from the best.}}
  \label{fig:conf}
  \Description{No visible trend wrt segment number, values range from
    .90 to .96}
  \rule[2ex]{\linewidth}{1pt} %
\end{figure}

And finally, same data again, ordered by correlation and with
95\% confidence intervals shown, in \autoref{fig:conf}.  This
certainly suggests that the best 5--10 segments are likely to be
better choices as proxies for the whole archive, and even more likely
that the \textit{worst} segments should be avoided.

Appendix \ref{rankings} gives a complete tabulation of segment rank
for all four years, using the three properties described
above (mime, language and length).

\subsubsection{Testing predictions across properties: Mime, 
  language and length}

We proceeded to see if the correlations computed as described in the
previous section can identify one or more segments from a Common Crawl
archive which are good proxies for the whole.

\begin{figure*}
  \includegraphics[width=\linewidth]{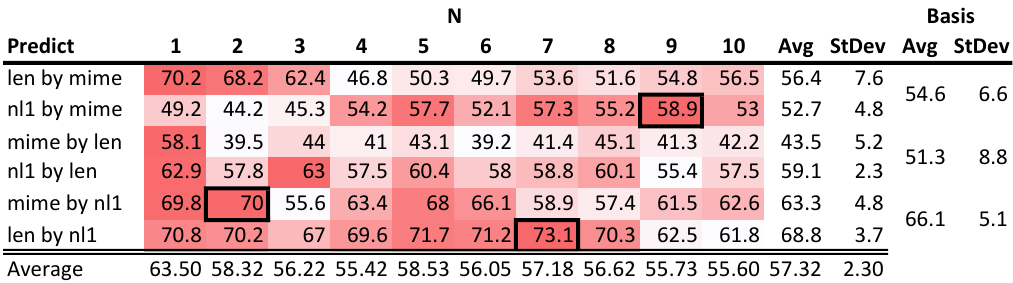}
  \caption[Heat map for prediction percentiles,
    2019-35]{Heat map for prediction percentiles,
    2019-35. \textnormal{The first column lists what property is being
      predicted by what: length (``len''), mime and language
      (``nl1'').  For example the
      first row gives the results for predicting what segment(s)
      should be used to model the distribution of the length property (the
      {\bf target})
      based on the best N segments from the distribution of the mime
      property (the {\bf basis}). The figure in a given cell is the
      percentile rank of the average of the segment-to-whole
      correlations from those segments in the target identified as the
      best N segments in the basis. The heat map colouring is
      per \textit{row}, that is, across all the predictions named in
      the first column as we vary the number used (N).  The
      black-margin cells highlight the best combination of basis and N
    for each target.  The final two columns give mean and standard
    deviation over all the predictions for a given basis, i.e. the
    based on the twenty cells to the left.}}
  \label{fig:heatmap1}
  \Description{The average rows and columns pretty well summarise the
  information from the heat map colouring:  N = 1 and 5 give the most
  consistently good results, and using the nl1 (language) property is better than
  the other two.  The latter is also underlined by the use of black margins
  around the best cell for each target:  for the mime target, the cell
[mime by nl1, 2] with value 70.0; for the len target, the cell [len by
nl1, 7] with value 73.1}  \rule[2ex]{\linewidth}{1pt} %
\end{figure*}

In particular, we can now determine what property, available from the
index, is best for predicting the segments to use for {\em other}
properties, as a prelude to working with properties which are not
available from the index.  \autoref{fig:heatmap1} shows, in the
form of heatmaps, all
possible pairings of three properties being used as the basis for
predicting one another as targets, computed as
described above in \autoref{proxy}.  For example
the cell in 5th column of the second row of the heatmap, with value
$57.7$, tests how well we can predict a good subset of the segments
from the 2019-35 archive to use to study the distribution of the
language property in that entire archive, using the identity of the top 5
segments from that archive in terms of the correlation between their
distribution of the \textit{mime} metric with its distribution in the
entire archive.

We can see that the predictions are not perfect, but we can learn what
we need from the table as a whole: The language property is both better on
average and more consistent. The length property is the least
reliable, although exactly \textit{why} it's so much worse at
predicting the mime property than the language property deserves
further investigation.

It's less clear what the best choice for N is.  Overall it seems that
almost any choice in the 1--5 range will be OK.  We'll look at that in
more detail in the next part, when we trial this approach in a case
where we really don't have all the target data available so we really
do need to use a proxy.

\section{Part 2: Using Last-Modified date to explore URI length}
\subsection{Methodology}
Common Crawl only started releasing archives in 2008, but we can still
get at least some insight even further back into the history of the
Web from even the most recent archives, by exploiting the information
in the Last-Modified HTTP response header.  Although this is optional,
it is
present in around 17\% of the successful responses for 2019-35. There are around 600 million responses with a
Last-Modified header out of around $3.4{\times}10^9$ in total.  The earliest
credible values archive are from the late 20th century.

According to the HTTP specification \cite{http}, the Last-Modified
header value must be in one of several standard textual formats,
preferably what is defined there as \textbf{HTTP-date}.  However not
all Web servers conform to that requirement.  We allowed a certain
limited amount of flexibility, for example in the (mis)placement of
``GMT'', but still ended up rejecting about .01\% as unusable as
written, and a further .1\% because they were not credible (too early
or in the future).

The accepted values were then converted into POSIX time format
\cite{time}, for easier sorting and comparison.

To complete our test of the proxying hypothesis, we wanted to use the
2023-40 archive, for which we did \textit{not} have the complete
archive, only the index, to start with.

\begin{figure*}
  \includegraphics[width=\linewidth]{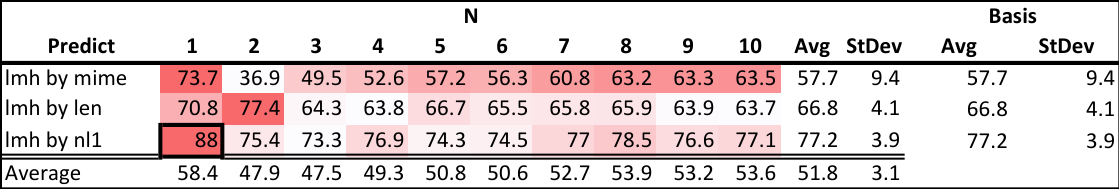}
  \caption[Heat map for prediction percentiles, 2023-40.]{Heat map for
    prediction percentiles, 2023-40. \textnormal{Last-Modified header
    count per year (``lmh'') predicted by mime, language and length.  Layout the same as
    \autoref{fig:heatmap1}. As there is only one target, the only the black border now is the best
    cell overall.}}
  \label{fig:heatmap2}
  \Description{Again the average rows and columns pretty well summarise the
  information from the heat map colouring:  N = 1 and 2 give the most
  consistently good results, and using the nl1 (language) property is better than
  the other two.  The black margin is just the best overall, namely
[lmh by nl1, 1] with value 88}
  \rule[2ex]{\linewidth}{1pt} %
\end{figure*}

Since we did have the full archive for 2019-35, we tested one further
property as target, namely the distribution of frequencies of
Last-Modified headers by year, to choose a proxy for our study of
2023-40.  \autoref{fig:heatmap2} shows the results of a further
round of testing in which the Last-Modified frequency was the target with the
same three properties as before as basis.

This shows somewhat less good results overall, but does confirm that
``language'' is the best predictor.  We chose to use $N = 2$ to reduce
the risk of hitting a bad segment by accident, and fetched just the
top two segments as measured by the ``language'' property from the
2023-40 index, that is, segments 56 and 12 (see Appendix
\ref{rankings}).

We extracted and processed the Last-Modified values from those two
2023-40 segments using the same method described above for 2019-35.
Taken together they contain around 12 million responses with a
Last-Modified header out of around 69 million in total, about 17\%,
the same ratio as observed for the 2019-35 data.  We then sorted and
tabulated the Last-Modified dates by year, by month within 2023 and by
day within September 2023.

\begin{figure}
  \centering
  \includegraphics[width=\linewidth]{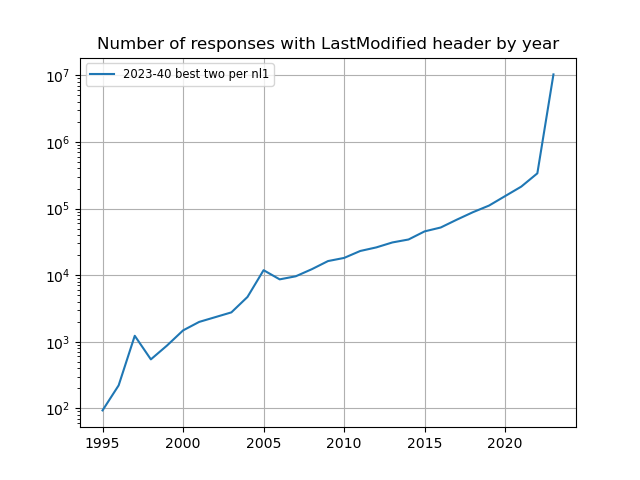}
  \caption[Last-Modified header counts by year
    ]{Last-Modified header counts by year.
     \textnormal{Based on uncorrected data from the 2023-40 archive. Semi-log plot for the y-axis.}}
  \label{fig:lmhcounts}
  \Description{Rising diagonal line from $10^2$ for 1995 with two
    upwards blips at 1997 and 2005, otherwise more-or-less linear up
    to $10^5.2$ for 2022, then steep spike to $10^7$ for 2023.}
  \rule[2ex]{\linewidth}{1pt} %
\end{figure}

\autoref{fig:lmhcounts} shows the number of headers found by year.
The years before 2000 are clearly very poorly represented, and are not
included in our subsequent analysis of URI length.  2005 also is badly
distorted from what looks like an anomaly in the Common Crawl data
gathering pipeline or the seeds from which it starts.  See Appendix
\ref{bogon} for details of the evidence that this is an anomaly and
the corrective action taken, which affects all data and figures after
\autoref{tab:seg_stats} and \autoref{fig:lmhcounts} and the discussion
thereof.

A semi-log plot is used in \autoref{fig:lmhcounts} because of the rapid
fall-off in pages with Last-Modified headers with values before 2023: over
10 million in 2023, only 1.3 million in total from earlier years.  More
details on this are provided in \autoref{web2}.

In order to look at URI length over time, we tabulated the overall length
of each URI from 2023-40 for which we have a Last-Modified date, as
well as the lengths of its parts (scheme, netloc, path, and query) and
three additional measures:
\begin{enumerate}
\item \textbf{idna}: whether or not a non-ascii netloc was present,
  encoded using punycode
\item \textbf{path\%}: for URIs with a non-empty path, the number of
  percent-encoded characters therein
\item \textbf{query\%}: for URIs with a non-empty query, the number of
  percent-encoded characters therein
\end{enumerate}

\subsection{Results}\label{web2}
\begin{figure}
  \centering
  \includegraphics[width=\linewidth]{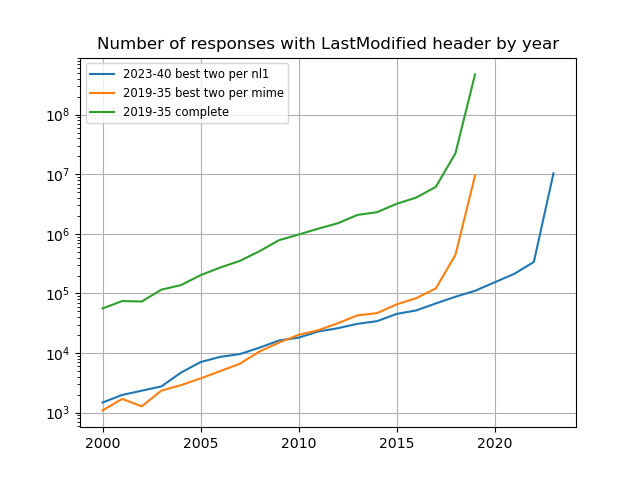}
  \caption[Comparing Last-Modified header counts by year
    ]{Comparing Last-Modified header counts by year.
     \textnormal{Based on data from the corrected 2019-35 and 2023-40
      archives. Semi-log plot for the y-axis.}}
  \label{fig:threeyears}
  \Description{Similar to \autoref{fig:lmhcounts} but starting at 2000 and
    adding including a similar lines for best_two_per_mime from
    2019-35 archive and the complete 2019-35 archive.  Using corrected
  data, so no blips. All three showing similar shape, but with 2019-35
  pair showing final spikes for 2019 to $10^7$ and nearly $10^9$
  respectively.  2023-40 and best_two from 2019-35 very close to one
  another from 2000 through 2017.}
  \rule[2ex]{\linewidth}{1pt} %
\end{figure}

\autoref{fig:threeyears} compares the proxied results from 2023-40
with both the full-archive results from 2019-35 and a comparable
best-two segment result from 2019-35.

This does suggest that the chosen proxy \textit{is} a representative
of the data as a whole.  Particularly reassuring is the conformance of
the 2023-40 proxy curve to the 2019-35 whole archive curve, as the
individual pages in those two years are almost all different: the
overlap between the two is less than $.4\%$ as measured by the Common
Crawl overlap statistics.

\subsubsection{URI lengths}

\begin{figure}
  \centering
  \includegraphics[width=\linewidth]{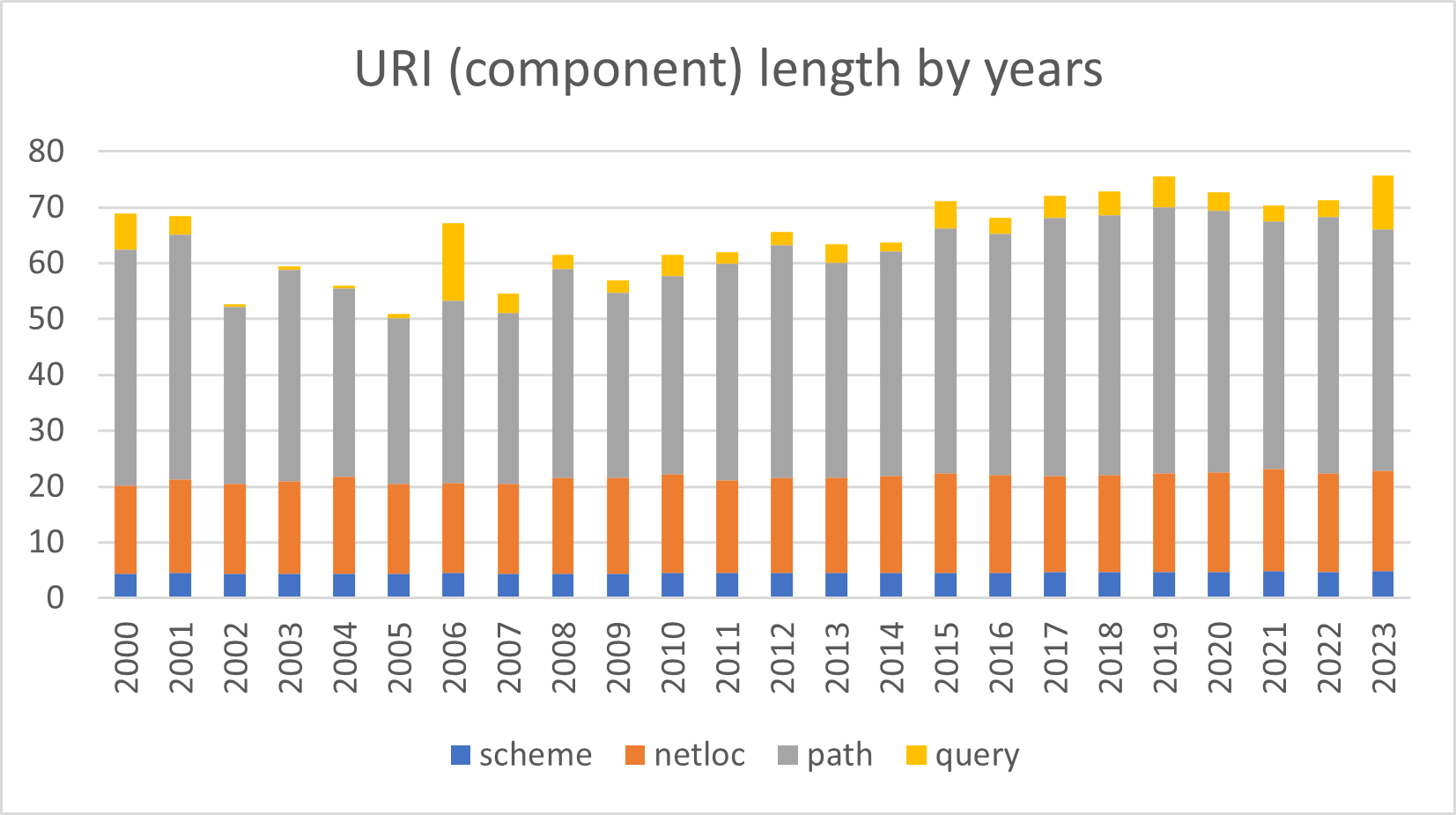}
  \caption[URI length, divided into component lengths, by year
    ]{URI length, divided into component lengths, by year.
     \textnormal{Based on data from the 2023-40
      archive.}}
  \label{fig:urilength}
  \Description{Scheme and netloc don't change, slow modest growth in
    path, query small and variable}
  \rule[2ex]{\linewidth}{1pt} %
\end{figure}

\begin{figure}
  \centering
  \includegraphics[width=\linewidth]{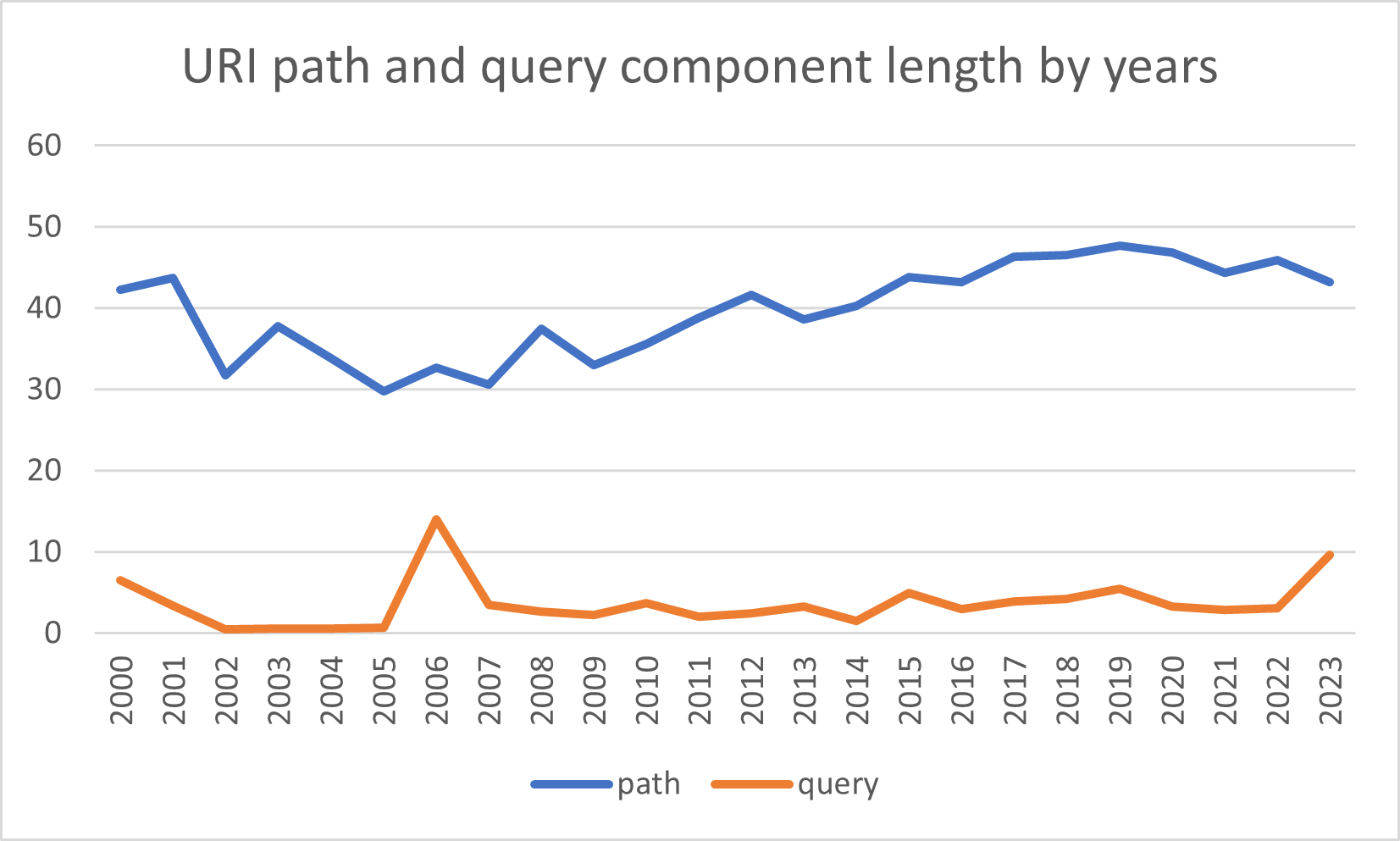}
  \caption[URI path and query length, by year
    ]{URI path and query length, by year.
     \textnormal{Based on data from the 2023-40
      archive.}}
  \label{fig:pathquery}
  \Description{Not much happening, upward blip for query in 2006}
  \rule[2ex]{\linewidth}{1pt} %
\end{figure}

The actual change in the length of URIs, and their component parts, is
shown in \autoref{fig:urilength}.  \autoref{fig:pathquery} shows
the change in the path and query components separately.  It appears
that our hypothesis was wrong, and that URI length is increasing only
slowly, and the change is more due to growth in path length than query
length.

\subsubsection{Last-minute Last-Modified values}

Something interesting emerges if we explore the Last-Modified values at a
detailed granularity.  We should have anticipated this given what we
know about the change from a human-authored web to a just-in-time
machine-generated web, and indeed some of our previous work \cite{lukasz2020} suggested
Common Crawl gives evidence of it.

\begin{figure}
  \centering
  \includegraphics[width=\linewidth]{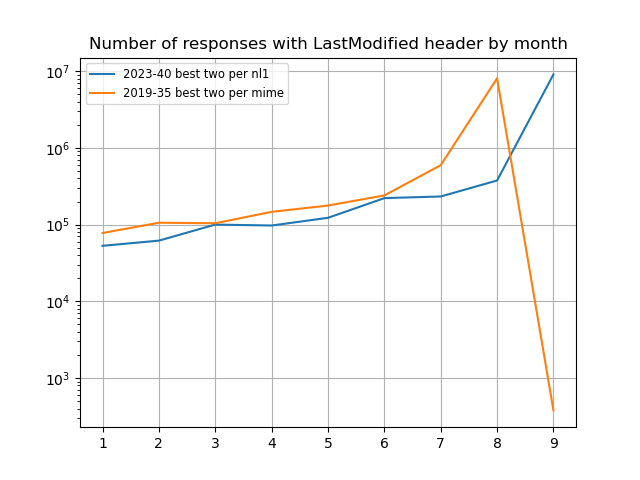}
  \caption[Last-Modified header counts by month
    ]{Last-Modified header counts by month.
     \textnormal{Based on data from the 2023-40
      archive. Semi-log plot for the y-axis.}}
  \label{fig:months}
  \Description{Two similar lines, with slow rise from $10^5$, then upward spike to 10^7 in
    crawl month}
  \rule[2ex]{\linewidth}{1pt} %
\end{figure}

\begin{figure}
  \centering
  \includegraphics[width=\linewidth]{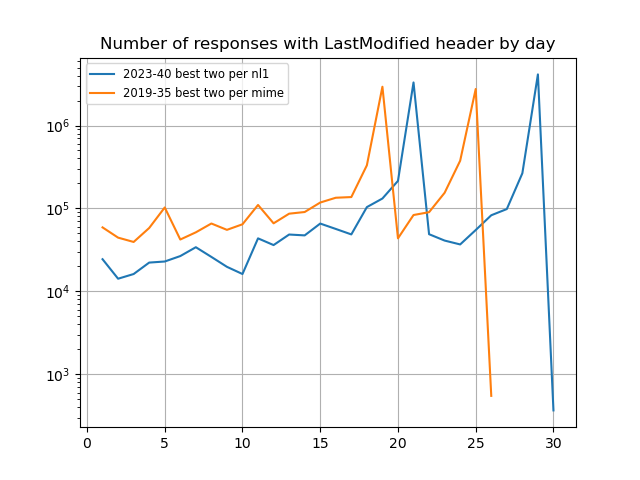}
  \caption[Last-Modified header counts by month
    ]{Last-Modified header counts by month.
     \textnormal{Based on data from the 2023-40
      archive. Semi-log plot for the y-axis.}}
  \label{fig:days}
  \Description{Two similar lines, a bit noisy, with slow rise from $10^4.4$, then upward spike to 10^6.3 in
    crawl month}
  \rule[2ex]{\linewidth}{1pt} %
\end{figure}

\renewcommand\figureautorefname{Figures}
\autoref{fig:months}
\renewcommand\figureautorefname{and}
\autoref{fig:days}
\renewcommand\figureautorefname{Figure}
focus in on
the year and month our proxy segments for 2023-40 were sampled.

The vast majority of the pages have Last-Modified dates on the two
days each of those proxy segments were crawled, 21 and 29 September 2023.  

\begin{figure}
  \centering
  \includegraphics[width=\linewidth]{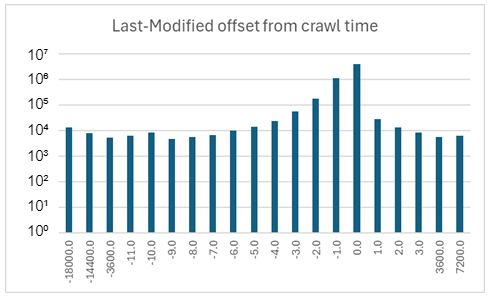}
  \caption[20 most frequent Last-Modified time offsets from crawl
  time, in seconds ]{20 most frequent Last-Modified time offsets from
    crawl time, in seconds.  \textnormal{Based on data from the
      2023-40 archive.  Overall N is 7,405,211, of which these account
      for 5,442,578 =~ 74\%. Negative column labels
      correspond to Last-Modified times \textit{before} their crawl time, positive
      ones to L-M times \textit{after}.  Semi-log plot for the y-axis.}}
  \label{fig:jit}
  \Description{All but -2, -1 and 0 offsets are below $10^5$, -1 at
    $10^6$ and 0 at $10^6.3$}
  \rule[2ex]{\linewidth}{1pt} %
\end{figure}

In order to check this further, we can compare the Last-Modified data
for pages crawled on those key days with the crawl timestamp, to see
how many of them were being created on the fly in response to
the crawler request.  \autoref{fig:jit} does exactly that, by
converting the crawl timestamps to 10-digit POSIX timestamps and
subtracting them from the Last-Modified times for all the pages for
which we have Last-Modified data that were crawled on 21 or 29 September.

53\% of the offsets are 0.0, 70\% are within three seconds of the
crawl time.  The marginal numbers are interesting in their own
right---they are at exactly -5, -4, -1 hours earlier and +1 and +2 hours later
than the crawl time.
That's what we would expect from just-in-time pages from various
likely areas if their Last-Modified values were
expressed as per their local timezone without being
converted to UTC and did not contain a timezone indication.  For
example, -14400, equivalent to 0 in UTC-4, is what we'd expect from
the East Coast of North America if local times in September were
reported without an explicit ``EDT'' (Daylight Savings
ended on 5 November in most of the USA).

\section{Discussion}
\subsection{Ranking segments as proxies for the whole}
For each of the four archives we've tested, we've shown that overall
the segmentation does a pretty good job of producing subsets which
closely resemble the whole archive with respect to the distribution of
media types: the {\em worst} rank correlation for a segment is .873,
and most are over 0.9.  We've done this in a way that can be easily
applied to other Common Crawl archives.  We plan to produce and
publish a list of segment rankings with respect to several properties
for every archive since January 2019, based on this approach.

We hope the availability of these rankings will in turn make it much
easier to carry out longitudinal studies using only 1\% or 10\% of a
series of full archives, bringing such work into reach for those
without access to large high-performance clusters.

\subsection{Explaining the growth in URI length}

The upward blip in the query in 2006 deserves further investigation,
but with $N$ still less than 10,000 it's probably just due to the lumpiness
of random processes:  after removing all the URIs from two
domains with queries which were sampled more than 100 times and whose
average query length is greater than 100, the mean query length is
down to a similar value to that of the surrounding years.

Once we paid more attention to the predominance of last-minute
Last-Modified values, it looks like the increase in overall size and
of the query component for 2023 is in fact from pages with very
small offsets between Last-Modified date and crawl time, compared
more `normal' offsets.  More investigation of more data is needed to
get a clear picture of what's going on.

\subsection{Last-minute Last-Modified values}

It is interesting to note that up until a few days before the crawl
date, the upward slope of the day-by-day curves in \autoref{fig:days}
is similar to that of the month-by-month curves \autoref{fig:months},
but slower that that of the year-by-year curves \autoref{fig:threeyears}.
It would be interesting to compare those slopes to whatever the
literature has to say about the actuarial profiles of human-authored
pages versus created-on-demand pages.  Or correlation with the amount
of JavaScript...


\begin{acks}
Access to the Cirrus UK National Tier-2 HPC Service at the Edinburgh
Parallel Computing Centre
(http://www.cirrus.ac.uk) used in this work was supported
by an EPSRC and UKRI HPC Access awards to Henry S. Thompson.

The work reported here has been stimulated by a number of MSc projects
over the last few years.  Particular thanks in the regard are due to
Jian Tong, Lukasz Domanski and Jingrui Chen.

Thanks to Sebastian Nagel of Common Crawl for many prompt and helpful
replies to many emails over the year, and to Greg Lindahl of Common
Crawl and Tom Morris for more recent help with consistency problems in the index
and the challenges of increasing load on the Common Crawl servers.

All use of Common Crawl archives and indices reported on in this work
conformed to the Common Crawl Terms of Use \cite{ccf3}.
\end{acks}

\bibliographystyle{ACM-Reference-Format}
\bibliography{wpaper}


\begin{thebibliography}{29}


\ifx \showCODEN    \undefined \def \showCODEN     #1{\unskip}     \fi
\ifx \showDOI      \undefined \def \showDOI       #1{#1}\fi
\ifx \showISBNx    \undefined \def \showISBNx     #1{\unskip}     \fi
\ifx \showISBNxiii \undefined \def \showISBNxiii  #1{\unskip}     \fi
\ifx \showISSN     \undefined \def \showISSN      #1{\unskip}     \fi
\ifx \showLCCN     \undefined \def \showLCCN      #1{\unskip}     \fi
\ifx \shownote     \undefined \def \shownote      #1{#1}          \fi
\ifx \showarticletitle \undefined \def \showarticletitle #1{#1}   \fi
\ifx \showURL      \undefined \def \showURL       {\relax}        \fi
\providecommand\bibfield[2]{#2}
\providecommand\bibinfo[2]{#2}
\providecommand\natexlab[1]{#1}
\providecommand\showeprint[2][]{arXiv:#2}

\bibitem[Amazon({[n.\,d.]})]%
        {AWSData}
\bibfield{author}{\bibinfo{person}{Amazon}.}
  \bibinfo{year}{[n.\,d.]}\natexlab{}.
\newblock \bibinfo{title}{Amazon Web Services (AWS) Open Data Sponsorship
  Program}.
\newblock \bibinfo{howpublished}{Web document}.
\newblock
\urldef\tempurl%
\url{https://aws.amazon.com/opendata/open-data-sponsorship-program/}
\showURL{%
\tempurl}
\newblock
\shownote{Retrieved: 22 November 2023}.


\bibitem[{Apache Software Foundation}({[n.\,d.]})]%
        {tika}
\bibfield{author}{\bibinfo{person}{{Apache Software Foundation}}.}
  \bibinfo{year}{[n.\,d.]}\natexlab{}.
\newblock \bibinfo{title}{Content analysis toolkit}.
\newblock \bibinfo{howpublished}{Web document}.
\newblock
\urldef\tempurl%
\url{https://tika.apache.org/}
\showURL{%
\tempurl}
\newblock
\shownote{Retrieved 7 December 2023}.


\bibitem[Archive({[n.\,d.]})]%
        {SURT}
\bibfield{author}{\bibinfo{person}{Internet Archive}.}
  \bibinfo{year}{[n.\,d.]}\natexlab{}.
\newblock \bibinfo{title}{Sort-friendly URI Reordering Transform}.
\newblock \bibinfo{howpublished}{Web document}.
\newblock
\urldef\tempurl%
\url{http://crawler.archive.org/articles/user_manual/glossary.html#surt}
\showURL{%
\tempurl}
\newblock
\shownote{Retrieved: 29 November 2022}.


\bibitem[Baack and Insights(2024)]%
        {moz2024}
\bibfield{author}{\bibinfo{person}{Stefan Baack} {and} \bibinfo{person}{Mozilla
  Insights}.} \bibinfo{year}{2024}\natexlab{}.
\newblock \bibinfo{title}{Training Data for the Price of a Sandwich}.
\newblock \bibinfo{howpublished}{Web document}.
\newblock
\urldef\tempurl%
\url{https://foundation.mozilla.org/en/research/library/generative-ai-training-data/common-crawl/}
\showURL{%
\tempurl}
\newblock
\shownote{Retrieved: 25 February 2024}.


\bibitem[Chapuis et~al\mbox{.}(2020)]%
        {Chap2020}
\bibfield{author}{\bibinfo{person}{Bertil Chapuis} {et~al\mbox{.}}}
  \bibinfo{year}{2020}\natexlab{}.
\newblock \showarticletitle{An Empirical Study of the Use of Integrity
  Verification Mechanisms for Web Subresources}. In
  \bibinfo{booktitle}{\emph{Proceedings of The Web Conference 2020}} (Taipei,
  Taiwan) \emph{(\bibinfo{series}{WWW '20})}. \bibinfo{publisher}{Association
  for Computing Machinery}, \bibinfo{address}{New York, NY, USA},
  \bibinfo{pages}{34–45}.
\newblock
\showISBNx{9781450370233}
\urldef\tempurl%
\url{https://doi.org/10.1145/3366423.3380092}
\showDOI{\tempurl}


\bibitem[Chen(2021)]%
        {cookies}
\bibfield{author}{\bibinfo{person}{Jingrui Chen}.}
  \bibinfo{year}{2021}\natexlab{}.
\newblock \bibinfo{booktitle}{\emph{A Survey on HTTP cookies: Do large Internet
  companies collect more information from users?}}
\newblock \bibinfo{type}{MSc dissertation}. \bibinfo{institution}{University of
  Edinburgh}.
\newblock


\bibitem[Chiniah et~al\mbox{.}(2019)]%
        {ChiniahEtAl2019}
\bibfield{author}{\bibinfo{person}{Aatish Chiniah}, \bibinfo{person}{Ayaz
  Chummun}, {and} \bibinfo{person}{Za Burkutallyïd}.}
  \bibinfo{year}{2019}\natexlab{}.
\newblock \showarticletitle{Categorising AWS Common Crawl Dataset using
  MapReduce}. In \bibinfo{booktitle}{\emph{2019 Conference on Next Generation
  Computing Applications (NextComp)}}. \bibinfo{pages}{1--6}.
\newblock
\urldef\tempurl%
\url{https://doi.org/10.1109/NEXTCOMP.2019.8883665}
\showDOI{\tempurl}


\bibitem[{Common Crawl}(2022)]%
        {ccf2}
\bibfield{author}{\bibinfo{person}{{Common Crawl}}.}
  \bibinfo{year}{2022}\natexlab{}.
\newblock \bibinfo{title}{{Common Crawl - Get Started}}.
\newblock \bibinfo{howpublished}{Web document}.
\newblock
\urldef\tempurl%
\url{https://commoncrawl.org/get-started}
\showURL{%
\tempurl}
\newblock
\shownote{Retrieved: 22 November 2023}.


\bibitem[{Common Crawl}(2023)]%
        {common_crawl}
\bibfield{author}{\bibinfo{person}{{Common Crawl}}.}
  \bibinfo{year}{2023}\natexlab{}.
\newblock \bibinfo{title}{Common Crawl - Open Repository of Web Crawl Data}.
\newblock \bibinfo{howpublished}{Web document}.
\newblock
\urldef\tempurl%
\url{https://commoncrawl.org}
\showURL{%
\tempurl}
\newblock
\shownote{Retrieved: 22 November 2023}.


\bibitem[{Common Crawl}(2024)]%
        {ccf3}
\bibfield{author}{\bibinfo{person}{{Common Crawl}}.}
  \bibinfo{year}{2024}\natexlab{}.
\newblock \bibinfo{title}{{Common Crawl - Terms of Use}}.
\newblock \bibinfo{howpublished}{Web document}.
\newblock
\urldef\tempurl%
\url{https://commoncrawl.org/terms-of-use}
\showURL{%
\tempurl}
\newblock
\shownote{Retrieved: 24 February 2024}.


\bibitem[Cox(2023)]%
        {atanh}
\bibfield{author}{\bibinfo{person}{Nick Cox}.}
  \bibinfo{year}{2011,2023}\natexlab{}.
\newblock \bibinfo{title}{How to calculate a confidence interval for Spearman's
  rank correlation}.
\newblock \bibinfo{howpublished}{Web document}.
\newblock
\urldef\tempurl%
\url{https://stats.stackexchange.com/a/18904}
\showURL{%
\tempurl}
\newblock
\shownote{Retrieved: 6 December 2023}.


\bibitem[Deutsch(1996)]%
        {gzip}
\bibfield{author}{\bibinfo{person}{Peter Deutsch}.}
  \bibinfo{year}{1996}\natexlab{}.
\newblock \bibinfo{booktitle}{\emph{GZIP file format specification version
  4.3}}.
\newblock \bibinfo{type}{Internet RFC}. \bibinfo{institution}{IETF}.
\newblock
\urldef\tempurl%
\url{https://www.ietf.org/rfc/rfc1952.html}
\showURL{%
\tempurl}
\newblock
\shownote{Retrieved: 22 November 2023}.


\bibitem[Domanski(2020)]%
        {lukasz2020}
\bibfield{author}{\bibinfo{person}{Lukasz Domanski}.}
  \bibinfo{year}{2020}\natexlab{}.
\newblock \bibinfo{booktitle}{\emph{Analysing Common Crawl - Efficient and
  Cost-Effective Processing of Large-Scale Data}}.
\newblock \bibinfo{type}{MSc dissertation}. \bibinfo{institution}{University of
  Edinburgh}.
\newblock


\bibitem[Du et~al\mbox{.}(2017)]%
        {DuEtAl2017}
\bibfield{author}{\bibinfo{person}{Yuheng Du} {et~al\mbox{.}}}
  \bibinfo{year}{2017}\natexlab{}.
\newblock \showarticletitle{Representativeness of latent dirichlet allocation
  topics estimated from data samples with application to common crawl}. In
  \bibinfo{booktitle}{\emph{2017 IEEE International Conference on Big Data (Big
  Data)}}. \bibinfo{pages}{1418--1427}.
\newblock
\urldef\tempurl%
\url{https://doi.org/10.1109/BigData.2017.8258075}
\showDOI{\tempurl}


\bibitem[Eberius et~al\mbox{.}(2015)]%
        {EberEtAl2015}
\bibfield{author}{\bibinfo{person}{Julian Eberius} {et~al\mbox{.}}}
  \bibinfo{year}{2015}\natexlab{}.
\newblock \showarticletitle{Building the Dresden Web Table Corpus: A
  Classification Approach}. In \bibinfo{booktitle}{\emph{2015 IEEE/ACM 2nd
  International Symposium on Big Data Computing (BDC)}}.
  \bibinfo{pages}{41--50}.
\newblock
\urldef\tempurl%
\url{https://doi.org/10.1109/BDC.2015.30}
\showDOI{\tempurl}


\bibitem[Fielding and Reschke(2014)]%
        {http}
\bibfield{author}{\bibinfo{person}{Roy~T. Fielding} {and}
  \bibinfo{person}{Julian Reschke}.} \bibinfo{year}{2014}\natexlab{}.
\newblock \bibinfo{title}{{Hypertext Transfer Protocol (HTTP/1.1): Conditional
  Requests}}.
\newblock \bibinfo{howpublished}{RFC 7232}.
\newblock
\urldef\tempurl%
\url{https://doi.org/10.17487/RFC7232}
\showDOI{\tempurl}
\newblock
\shownote{Retrieved: 22 November 2023}.


\bibitem[Hantke and Stock(2022)]%
        {FandS2022}
\bibfield{author}{\bibinfo{person}{Florian Hantke} {and} \bibinfo{person}{Ben
  Stock}.} \bibinfo{year}{2022}\natexlab{}.
\newblock \showarticletitle{HTML violations and where to find them: a
  longitudinal analysis of specification violations in HTML}. In
  \bibinfo{booktitle}{\emph{Proceedings of the 22nd ACM Internet Measurement
  Conference}} (Nice, France) \emph{(\bibinfo{series}{IMC '22})}.
  \bibinfo{publisher}{Association for Computing Machinery},
  \bibinfo{address}{New York, NY, USA}, \bibinfo{pages}{358–373}.
\newblock
\showISBNx{9781450392594}
\urldef\tempurl%
\url{https://doi.org/10.1145/3517745.3561437}
\showDOI{\tempurl}


\bibitem[{{IEEE}}(2017)]%
        {time}
\bibfield{author}{\bibinfo{person}{{{IEEE}}}.} \bibinfo{year}{2017}\natexlab{}.
\newblock \bibinfo{title}{The Open Group Base Specifications Issue 7, 2018
  edition \\ IEEE Std 1003.1-2017: POSIX.1a}.
\newblock \bibinfo{howpublished}{Web document}.
\newblock
\urldef\tempurl%
\url{https://pubs.opengroup.org/onlinepubs/9699919799/basedefs/V1_chap04.html#tag_04_16}
\showURL{%
\tempurl}
\newblock
\shownote{Retrieved: 25 November 2023}.


\bibitem[{International Internet Preservation Consortium}(2017)]%
        {warc}
\bibfield{author}{\bibinfo{person}{{International Internet Preservation
  Consortium}}.} \bibinfo{year}{2017}\natexlab{}.
\newblock \bibinfo{howpublished}{Web document}.
\newblock
\urldef\tempurl%
\url{https://iipc.github.io/warc-specifications/specifications/warc-format/warc-1.1/}
\showURL{%
\tempurl}
\newblock
\shownote{Retrieved: 29 November 2022}.


\bibitem[Kreymer(2015a)]%
        {cdx}
\bibfield{author}{\bibinfo{person}{Ilya Kreymer}.}
  \bibinfo{year}{2015}\natexlab{a}.
\newblock \bibinfo{howpublished}{Web document}.
\newblock
\urldef\tempurl%
\url{https://github.com/ikreymer/pywb/wiki/CDX-Index-Format#zipnum-sharded-cdx}
\showURL{%
\tempurl}
\newblock
\shownote{Retrieved: 22 November 2023}.


\bibitem[Kreymer(2015b)]%
        {cci}
\bibfield{author}{\bibinfo{person}{Ilya Kreymer}.}
  \bibinfo{year}{2015}\natexlab{b}.
\newblock \bibinfo{title}{Announcing the Common Crawl Index}.
\newblock \bibinfo{howpublished}{Web document}.
\newblock
\urldef\tempurl%
\url{https://commoncrawl.org/blog/announcing-the-common-crawl-index}
\showURL{%
\tempurl}
\newblock
\shownote{Retrieved: 22 November 2023}.


\bibitem[Luccioni and Viviano(2021)]%
        {LandV2021}
\bibfield{author}{\bibinfo{person}{Alexandra Luccioni} {and}
  \bibinfo{person}{Joseph Viviano}.} \bibinfo{year}{2021}\natexlab{}.
\newblock \showarticletitle{What{'}s in the Box? An Analysis of Undesirable
  Content in the {C}ommon {C}rawl Corpus}. In
  \bibinfo{booktitle}{\emph{Proceedings of the 59th Annual Meeting of the
  Association for Computational Linguistics and the 11th International Joint
  Conference on Natural Language Processing (Volume 2: Short Papers)}},
  \bibfield{editor}{\bibinfo{person}{Chengqing Zong}, \bibinfo{person}{Fei
  Xia}, \bibinfo{person}{Wenjie Li}, {and} \bibinfo{person}{Roberto Navigli}}
  (Eds.). \bibinfo{publisher}{Association for Computational Linguistics},
  \bibinfo{address}{Online}, \bibinfo{pages}{182--189}.
\newblock
\urldef\tempurl%
\url{https://doi.org/10.18653/v1/2021.acl-short.24}
\showDOI{\tempurl}


\bibitem[Maler et~al\mbox{.}(2008)]%
        {xml}
\bibfield{author}{\bibinfo{person}{Eve Maler}, \bibinfo{person}{Jean Paoli},
  {et~al\mbox{.}}} \bibinfo{year}{2008}\natexlab{}.
\newblock \bibinfo{booktitle}{\emph{Extensible Markup Language (XML) 1.0 (Fifth
  Edition)}}.
\newblock \bibinfo{type}{Fifth Edition of a W3C Recommendation}.
  \bibinfo{institution}{W3C}.
\newblock
\urldef\tempurl%
\url{http://www.w3.org/TR/xml/}
\showURL{%
\tempurl}
\newblock
\shownote{Retrieved: 29 November 2022}.


\bibitem[Nagel(2022)]%
        {ccf1}
\bibfield{author}{\bibinfo{person}{Sebastian Nagel}.}
  \bibinfo{year}{2022}\natexlab{}.
\newblock \bibinfo{title}{{Common Crawl (Getting Started)}}.
\newblock \bibinfo{howpublished}{Web document}.
\newblock
\urldef\tempurl%
\url{https://commoncrawl.org/the-data/get-started/}
\showURL{%
\tempurl}
\newblock
\shownote{Retrieved: 29 November 2022}.


\bibitem[Panchenko et~al\mbox{.}(2018)]%
        {PanchEtAl2018}
\bibfield{author}{\bibinfo{person}{Alexander Panchenko} {et~al\mbox{.}}}
  \bibinfo{year}{2018}\natexlab{}.
\newblock \showarticletitle{{Building a Web-Scale Dependency-Parsed Corpus from
  CommonCrawl}}. In \bibinfo{booktitle}{\emph{Proceedings of the Eleventh
  International Conference on Language Resources and Evaluation (LREC 2018)}},
  \bibfield{editor}{\bibinfo{person}{Nicoletta Calzolari} {et~al\mbox{.}}}
  (Eds.). \bibinfo{publisher}{European Language Resources Association (ELRA)},
  \bibinfo{address}{Miyazaki, Japan}.
\newblock
\showISBNx{979-10-95546-00-9}


\bibitem[Project({[n.\,d.]})]%
        {statm}
\bibfield{author}{\bibinfo{person}{Statsmodels Project}.}
  \bibinfo{year}{[n.\,d.]}\natexlab{}.
\newblock \bibinfo{title}{statistical models, hypothesis tests, and data
  exploration}.
\newblock \bibinfo{howpublished}{Web document}.
\newblock
\urldef\tempurl%
\url{https://www.statsmodels.org/stable/index.html}
\showURL{%
\tempurl}
\newblock
\shownote{Retrieved: 28 November 2022}.


\bibitem[Sites(2013)]%
        {cld2}
\bibfield{author}{\bibinfo{person}{Dick Sites}.}
  \bibinfo{year}{2013}\natexlab{}.
\newblock \bibinfo{title}{Compact Language Detector 2}.
\newblock \bibinfo{howpublished}{Web document}.
\newblock
\urldef\tempurl%
\url{https://github.com/CLD2Owners/cld2}
\showURL{%
\tempurl}
\newblock
\shownote{Retrieved 7 December 2023}.


\bibitem[Thompson and Tong(2018)]%
        {TandT2018}
\bibfield{author}{\bibinfo{person}{Henry~S. Thompson} {and}
  \bibinfo{person}{Jian Tong}.} \bibinfo{year}{2018}\natexlab{}.
\newblock \showarticletitle{Can Common Crawl Reliably Track Persistent
  Identifier (PID) Use Over Time}. In \bibinfo{booktitle}{\emph{Companion
  Proceedings of the The Web Conference 2018}} (Lyon, France)
  \emph{(\bibinfo{series}{WWW '18})}. \bibinfo{publisher}{International World
  Wide Web Conferences Steering Committee}, \bibinfo{address}{Republic and
  Canton of Geneva, CHE}, \bibinfo{pages}{1749–1755}.
\newblock
\showISBNx{9781450356404}
\urldef\tempurl%
\url{https://doi.org/10.1145/3184558.3191636}
\showDOI{\tempurl}


\bibitem[Virtanen et~al\mbox{.}(2020)]%
        {scipy}
\bibfield{author}{\bibinfo{person}{Pauli Virtanen} {et~al\mbox{.}}}
  \bibinfo{year}{2020}\natexlab{}.
\newblock \showarticletitle{{{SciPy} 1.0: Fundamental Algorithms for Scientific
  Computing in Python}}.
\newblock \bibinfo{journal}{\emph{Nature Methods}}  \bibinfo{volume}{17}
  (\bibinfo{year}{2020}), \bibinfo{pages}{261--272}.
\newblock
\urldef\tempurl%
\url{https://doi.org/10.1038/s41592-019-0686-2}
\showDOI{\tempurl}


\end{thebibliography}
\appendix
\section{Detection and correction of Last-Modified issue}\label{bogon}

As noted above in the discussion of \autoref{fig:lmhcounts}, the
frequency of Last-Modified dates in 2005 stands out from the
surrounding years.  \autoref{tab:anomalya} shows the details of this
in increasingly fine granularity, to the surprising point where we see
that the whole difference is down to a single day.

Further inspection of the Last-Modified values for that day revealed
that the huge increase on that day was due to a
single Last-Modified value, namely \texttt{Sun, 24 Apr 2005 04:29:37 GMT}.

\begin{table}
    \caption[Evidence
      for anomaly in Last-Modified frequency]{Evidence
      for anomaly in Last-Modified frequency
\textnormal{By year, month and day.  The column in bold is the
  anomaly.  Based on data from the uncorrected 2019-35
      archive.}}
  \label{tab:anomalya}
  \begin{tabular}{r|rrrrr}
   Source&\multicolumn{5}{c}{counts}\\
   \midrule
   \multirow {2}{*}{Years}&2003&2004&\textbf{2005}&2006&2007\\
&115926&138295&\textbf{567693}&272565&351429\\
   \multirow {2}{*}{Months in 2005}&Feb&Mar&\textbf{Apr}&May&Jun\\
&17114&15083&\textbf{378061}&14640&19119\\
   \multirow {2}{*}{Days in 2005-04}&22&23&\textbf{24}&25&26\\
&362&215&\textbf{365113}&1167&554
  \end{tabular}
\end{table}

\begin{figure}
  \centering
  \includegraphics[width=\linewidth]{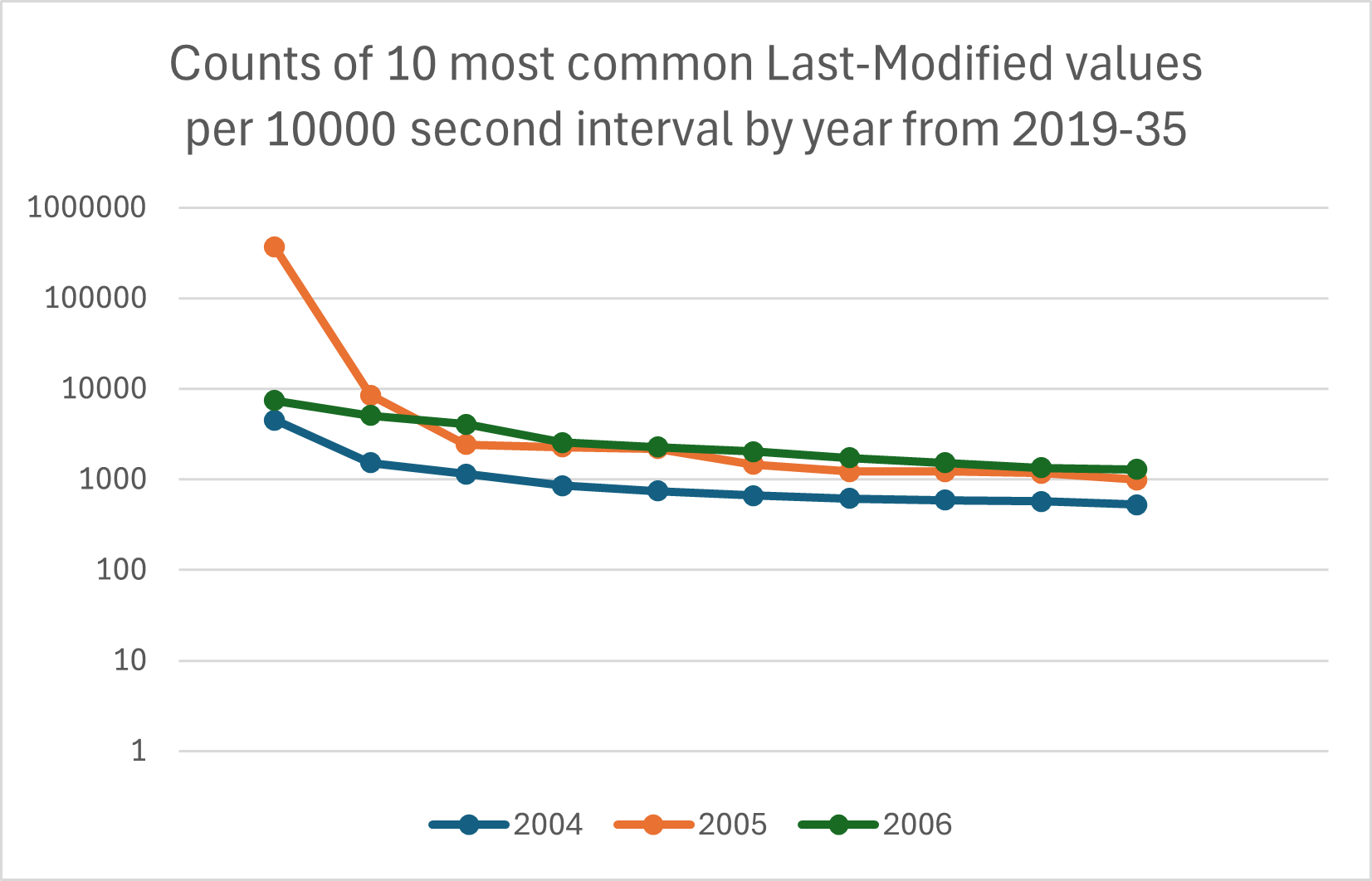}
  \caption[Last-Modified frequency by 10000 second
  ranges]{Last-Modified frequency by 10000 second ranges
\textnormal{The left-most orange point is for the range including
  1114316977. Based on data from the uncorrected 2019-35
      archive. Semi-log plot for the y-axis.}}
  \label{fig:bogon}
  \Description{Mostly flat between $10^3$ and $10^4$, with big spike up to $10^5.5$ for most common 2005}
  \rule[2ex]{\linewidth}{1pt} %
\end{figure}

We confirmed the anomalous nature of anything like this by tabulating
the frequency of all the Last-Modified values found in the 2019-35 archive in 10000-second intervals,
by counting the first 6 digits of the corresponding 10-digit POSIX
time value.  \autoref{fig:bogon} shows the 10 most frequent counts for any
such interval in 2005 and similar values for the surrounding years,
that is, not necessarily for the same interval, but the
\textit{same-ranked} interval.  The outlier, by
nearly two decimal orders of magnitude, is the interval containing 1114316977, the
10-digit POSIX version of the anomalous value.

\begin{table}
    \caption[Zooming in on anomaly in Last-Modified counts]{Zooming in on anomaly in Last-Modified counts
\textnormal{Top row for each archive shows counts for the most common
  interval for Last-Modified values, bottom
  row for the second-most common.  The figures in bold are for the
  interval containing 1114316977.  Based on data from the uncorrected 2019-35
      archive.}}
  \label{tab:anomalyb}
  \begin{tabular}{r|rrr}
&\multicolumn{3}{c}{Last-Modified year}\\
Archive&2004&2005&2006 \\
\midrule
\multirow {2}{*}{2019-35 (whole)}&4511&\textbf{364934}&7400 \\
&1521&8521&5047 \\\cline{2-4}
\multirow {2}{*}{2023-40 (6 segs only)}&215&\textbf{13408}&857 \\
&187&824&556
    \end{tabular}
\end{table}

\autoref{tab:anomalyb} zooms in on counts for full 10-digit Last-Modified
values from the year in question, 2005, and one year on either side.

This unique value accounts for all but 9 cases of its corresponding
6-figure value for 2019-35 and all but 3 for 2023-40.  This 10-figure
value never appears in the other two years in either archive.  This
level of frequency for a single exact Last-Modified value is
unprecedented.  The next most common full 10-digit LM in the 6-segment
2019-35 archive data for these three years occurs only 7329 times, in
the 6 segments of 2023-40 only 805 times, so we're looking at factors
of 49 and 15.

On this basis I judged that the twin coincidences of a super-unlikely
over-count for a single LM date, across over 900 domains in 2023-40,
over 31000 in 2019-35 (mostly .ru,
.ua, .su, .am, .kz,
.xm--p1ai == {\cyr rf}), 
and the fact that it occurs in \textit{different} archives from mostly
\textit{different} domains (more than half the domains from 2023-40
don't show up again in 2019-35) is an indication of a problem somewhere.

The subsequent analyses reported here are therefore based on
Last-Modified data accordingly are based on data with all entries with
the 1114316977 Last-Modified value removed, amounting to 378,330 out
of approximately $521{\times}10^6$.

I'm working with the Common Crawl technical staff to try to find the
cause, but as of this writing without success.
\section{Segment ranking tables}\label{rankings}

\begin{table}[b]
  \label{tab:ranks}
    \caption{Segment ranks (best-to-worst) based on media type distribution
    correlations}
  \begin{tabular}{r|rrrr}
    &\multicolumn{3}{c}{Archive}\\
    rank&2019-35&2020-34&2021-31 \\
    \midrule
1&33&79&49\\
2&23&71&83\\
3&10&34&87\\
4&34&83&81\\
5&28&73&62\\
6&69&38&93\\
7&94&20&77\\
8&29&88&86\\
9&9&44&94\\
10&26&45&99
    \end{tabular}
\end{table}
The accompanying table gives top-10 complete segment rankings, sorted on
correlation of the mime property discussed above
in \autoref{represent} and divided into
deciles, for the first three years reported here.  Full rankings, as
well as the index for 2019-35 with Last-Modified times added, will be
published as soon as I can find a free place to host them.

\end{document}